\documentclass[aps,pra,twocolumn,showpacs,superscriptaddress]{revtex4}

\usepackage{amsmath}
\usepackage{amssymb}

\begin{document}
\title{Addendum to ``Experimental demonstration of a quantum protocol
for Byzantine agreement and liar detection''}
\author{Sascha Gaertner}
\affiliation{Max-Planck-Institut f\"{u}r
Quantenoptik, D-85748 Garching, Germany} \affiliation{Fakult\"{a}t
f\"{u}r Physik, Ludwig-Maximilians-Universit\"{a}t, D-80799
M\"{u}nchen, Germany}
\author{Mohamed Bourennane}
\affiliation{Department of Physics, Stockholm University, SE-10691
Stockholm, Sweden}
\author{Christian Kurtsiefer}
\affiliation{Department of Physics, National University of
Singapore, 117542 Singapore, Singapore}
\author{Ad\'an Cabello}
\email{adan@us.es} \affiliation{Departamento de F\'{\i}sica Aplicada
II, Universidad de Sevilla, E-41012 Sevilla, Spain}
\author{Harald Weinfurter}
\affiliation{Max-Planck-Institut f\"{u}r Quantenoptik, D-85748
Garching, Germany} \affiliation{Fakult\"{a}t f\"{u}r Physik,
Ludwig-Maximilians-Universit\"{a}t, D-80799 M\"{u}nchen, Germany}

\date{\today}


\begin{abstract}
Gao {\em et al.} [Phys. Rev. Lett. {\bf 101}, 208901 (2008)] have
described a possible intercept-resend attack for the quantum
protocol for detectable Byzantine agreement in Phys. Rev. Lett. {\bf
100}, 070504 (2008). Here we describe an extension of the protocol
which defeats such attacks.
\end{abstract}



\maketitle


Recently \cite{GGWZ08}, Gao {\em et al.} have pointed out that there
exists an intercept-resend attack for the quantum protocol for
detectable Byzantine agreement described in \cite{GBKCW08}. Here we
describe how such attacks alter the entanglement of the distributed
multipartite state and thus can be revealed by further analysis of
the acquired data.

The protocol uses the four-photon state
\begin{eqnarray}
|\Psi^{(4)}\rangle_{abcd} & = & {\frac{1}{2 \sqrt{3}}}
(2|0011\rangle-|0101\rangle-|0110\rangle-|1001\rangle \nonumber \\
& & -|1010\rangle+2|1100\rangle)_{abcd}, \label{state}
\end{eqnarray}
to distribute lists securely to three parties which then enable them
to achieve the detectable Byzantine agreement.

In \cite{GBKCW08}, step {\em (iii)} stated: ``$C$ randomly chooses a
position from his list and asks $A$ and $B$ to inform him about
their results on the same position. If all parties have measured in
the same basis, their results must be suitably correlated.'' In the
attack proposed in \cite{GGWZ08}, the traitor intercepts the qubits
sent to one of the loyal generals and performs measurements on them.
These measurements are single-qubit measurements in the
$\{|0\rangle,|1\rangle\}$ basis of eigenstates of the Pauli matrix
$\sigma_z$ or in the $\{|\bar{0}\rangle, |\bar{1}\rangle\}$ basis
[where $|\bar{0\rangle} = (|0\rangle + |1\rangle)/\sqrt{2}$ and
$|\bar{1}\rangle = (|0\rangle - |1\rangle)/\sqrt{2}$] of eigenstates
of the Pauli matrix $\sigma_x$. Then, the traitor resends the qubits
in the state resulting from the previous measurements to the loyal
general. With this method, the traitor can obtain the other
generals' secret lists which he can use to mislead the loyal
generals into following different plans.

Since the roles of $B$ and $C$ in the protocol in \cite{GBKCW08} are
symmetrical, there are only three cases to be dealt with:

$(I)$ $A$ (the commanding general) is the traitor and intercepts,
measures, and resends the qubit sent to $B$ (qubit $c$).

$(II)$ $B$ is the traitor and intercepts, measures, and resends the
two qubits sent to $A$ (qubits $a$ and $b$).

$(III)$ $B$ is the traitor and intercepts, measures, and resends the
qubit sent to $C$ (qubit $d$).

These attacks indeed do not alter the perfect correlations utilized
in the protocol so far. However, they obviously change the
entanglement of the four photon state and thus the correlations
between the results if the parties measured along different
directions. This can be used to detect the attack. There are several
ways to foil these attacks. Here we present a simple method which
does not require additional measurements and in which the traitor
does not need to announce his actual results. In the following, we
present an example of the method for each of the three cases.

$(I)$ Suppose that $A$ is the traitor and intercepts and measures
the qubit sent to $B$ in the basis of eigenstates of $\sigma_z$. In
half of the cases, $A$ obtains the result corresponding to
$|0\rangle_c$ and resends the state $|0\rangle_c$ to $B$. Therefore,
in these cases the state of the four qubits is
\begin{equation}
|\Psi'\rangle_{abcd} = \frac{1}{\sqrt{6}}
(-|0101\rangle-|1001\rangle +2|1100\rangle)_{abcd}.
\end{equation}
In the other half of the cases, $A$ obtains the result corresponding
to $|1\rangle_c$ and resends $B$ the state $|1\rangle_c$. Therefore,
in these cases the state of the four qubits is
\begin{equation}
|\Psi''\rangle_{abcd} = \frac{1}{\sqrt{6}}
(2|0011\rangle-|0110\rangle-|1010\rangle)_{abcd}.
\end{equation}
Therefore, after $A$'s attack, the state of the four qubits is the
mixed state
\begin{equation}
\rho_{abcd}=\frac{1}{2}(|\Psi'\rangle
\langle\Psi'|+|\Psi''\rangle \langle\Psi''|)_{abcd}.
\end{equation}
Suppose that $B$ and $C$ measure their qubits in the basis of
eigenstates of $\sigma_x$. Then, $C$ asks $A$ and $B$ for their
results. When $C$ checks them, $C$ expects that $\langle
\sigma_x^{(c)} \otimes \sigma_x^{(d)} \rangle_{\Psi^{(4)}} =
\frac{1}{3}$, which is the prediction for the state
$|\Psi^{(4)}\rangle_{abcd}$. However, for the mixed state
$\rho_{abcd}$, $C$ finds $\langle \sigma_x^{(c)} \otimes
\sigma_x^{(d)} \rangle_{\rho_{abcd}} = 0$.

$(II)$ If $B$ is the traitor and intercepts and measures qubits $a$
and $b$ in the basis of eigenstates of $\sigma_z$ (resulting in some
mixed state $\rho'_{abcd}$), and $A$ measures qubits $a$ and $b$ in
the basis of eigenstates of $\sigma_x$, then $C$ expects that
$\langle \sigma_x^{(a)} \otimes \sigma_x^{(b)} \rangle =
\frac{1}{3}$ but finds $\langle \sigma_x^{(a)} \otimes
\sigma_x^{(b)} \rangle_{\rho'_{abcd}} = 0$.

$(III)$ If $B$ is the traitor and intercepts and measures qubit $d$
in the basis of eigenstates of $\sigma_z$, and $A$ and $C$ measure
their qubits in the basis of eigenstates of $\sigma_x$, then $C$
expects that $\langle \sigma_x^{(a)} \otimes \sigma_x^{(d)}
\rangle_{\Psi^{(4)}} = \langle \sigma_x^{(b)} \otimes \sigma_x^{(d)}
\rangle_{\Psi^{(4)}} = -\frac{2}{3}$ but finds $\langle
\sigma_x^{(a)} \otimes \sigma_x^{(d)} \rangle_{\rho''_{abcd}} =
\langle \sigma_x^{(b)} \otimes \sigma_x^{(d)}
\rangle_{\rho''_{abcd}} = 0$.

If $C$ (or the other generals, since they exchange their roles in
the next step) obtains results that do not satisfy the quantum
predictions for the state $|\Psi^{(4)}\rangle_{abcd}$, then the
loyal generals decide to abort the protocol; otherwise, the loyal
generals can reach an agreement using the lists.


\begin{table}[t]
\caption{\label{tab2} Experimental data showing that an
intercept-resend attack \cite{GGWZ08} has not occurred in the
experiment described in \cite{GBKCW08}.}
\begin{ruledtabular}
{\begin{tabular}{ccc}
 & Experimental result & Result for $|\Psi^{(4)}\rangle_{abcd}$ \\
\hline $\langle \sigma_x^{(a)} \otimes \sigma_x^{(b)}
\rangle$ & $0.262 \pm 0.025$ & $1/3$ \\
$\langle \sigma_z^{(a)} \otimes \sigma_z^{(b)}
\rangle$ & $0.273 \pm 0.025$ & $1/3$ \\
$\langle \sigma_x^{(a)} \otimes \sigma_x^{(c)}
\rangle$ & $-0.602 \pm 0.021$ & $-2/3$ \\
$\langle \sigma_z^{(a)} \otimes \sigma_z^{(c)}
\rangle$ & $-0.631 \pm 0.02$ & $-2/3$ \\
$\langle \sigma_x^{(a)} \otimes \sigma_x^{(d)}
\rangle$ & $-0.612 \pm 0.02$ & $-2/3$ \\
$\langle \sigma_z^{(a)} \otimes \sigma_z^{(d)}
\rangle$ & $-0.663 \pm 0.019$ & $-2/3$ \\
$\langle \sigma_x^{(b)} \otimes \sigma_x^{(c)}
\rangle$ & $-0.603 \pm 0.021$ & $-2/3$ \\
$\langle \sigma_z^{(b)} \otimes \sigma_z^{(c)}
\rangle$ & $-0.615 \pm 0.02$ & $-2/3$ \\
$\langle \sigma_x^{(b)} \otimes \sigma_x^{(d)}
\rangle$ & $-0.61 \pm 0.02$ & $-2/3$ \\
$\langle \sigma_z^{(b)} \otimes \sigma_z^{(d)}
\rangle$ & $-0.621 \pm 0.02$ & $-2/3$ \\
$\langle \sigma_x^{(c)} \otimes \sigma_x^{(d)}
\rangle$ & $0.334 \pm 0.025$ & $1/3$ \\
$\langle \sigma_z^{(c)} \otimes \sigma_z^{(d)}
\rangle$ & $0.326 \pm 0.024$ & $1/3$ \\
\end{tabular}}
\end{ruledtabular}
\end{table}


We show that in the experiment described in \cite{GBKCW08}, actual
experimental data prove that the attack \cite{GGWZ08} has not
occurred. Table \ref{tab2} contains all the tests described above
and those obtained from them by changing $\sigma_x^{(i)}$ by
$\sigma_z^{(i)}$ for all qubits, exploiting the fact that the state
(\ref{state}) is invariant under the same unitary transformation
applied to the four qubits.


In conclusion, there is a simple method that makes the protocol in
\cite{GBKCW08} secure against the attack proposed in \cite{GGWZ08}.




\begin{thebibliography}{20}

\bibitem{GGWZ08}
F. Gao, F.-Z. Guo, Q.-Y. Wen, and F.-C. Zhu,
Phys. Rev. Lett. {\bf 101}, 208901 (2008).

\bibitem{GBKCW08}
S. Gaertner, M. Bourennane, C. Kurtsiefer, A. Cabello, and H.
Weinfurter,
Phys. Rev. Lett. {\bf 100}, 070504 (2008).

\end{thebibliography}
\end{document}